**The Role of Open-Source LLMs in Shaping the Future of GeoAI**

Xiao Huang[1], Zhengzhong Tu[2], Xinyue Ye[3,*], Michael Goodchild[4]

[1]Department of Environmental Sciences, Emory University, Atlanta, GA 30322, USA; xiao.huang2@emory.edu

[2]Department of Computer Science and Engineering, Texas A&M University, College Station, TX 77840, USA; tzz@tamu.edu

[3]Department of Landscape Architecture and Urban Planning & Center for Geospatial Sciences, Applications and Technology, Texas A&M University, College Station, TX 77840, USA; xinyue.ye@tamu.edu (corresponding author)

[4]Department of Geography, University of California, Santa Barbara, CA 93106, USA; good@geog.ucsb.edu

**Abstract:**

Large Language Models (LLMs) are transforming geospatial artificial intelligence (GeoAI), offering new capabilities in data processing, spatial analysis, and decision support. This paper examines the open-source paradigm's critical role in this transformation. While proprietary LLMs offer accessibility, they often limit the customization, interoperability, and transparency vital for specialized geospatial tasks. Conversely, open-source alternatives significantly advance Geographic Information Science (GIScience) by fostering greater adaptability, reproducibility, and community-driven innovation. Open frameworks empower researchers to tailor solutions, integrate cutting-edge methodologies (e.g., reinforcement learning, advanced spatial indexing), and align with FAIR (Findable, Accessible, Interoperable, and Reusable) principles. However, the growing reliance on any LLM necessitates careful consideration of security vulnerabilities, ethical risks, and robust governance for AI-generated geospatial outputs. This paper argues that GIScience advances best not through a single model type, but by cultivating a diverse, interoperable ecosystem combining open-source foundations for innovation, custom geospatial models, and interdisciplinary collaboration. By critically evaluating the opportunities and challenges of open-source LLMs within the broader GeoAI landscape, this work contributes to a thorough discourse on leveraging LLMs to effectively advance spatial research, policy, and decision-making in an equitable, sustainable, and scientifically rigorous manner.

## 1. Introduction

Open science has emerged as a powerful driver of innovation across many fields, including Geographic Information Science (GIScience). Researchers and practitioners have increasingly recognized the value of transparent, community-driven platforms for geospatial analysis and decision-making, particularly as traditional, closed-source frameworks can constrain customization and limit collaborative knowledge-building. In recent years, rapid developments in generative artificial intelligence (GenAI) have further underscored the importance of open-source alternatives (Wang et al., 2024). Large Language Models (LLMs) such as ChatGPT (OpenAI, 2023) and other proprietary offerings have demonstrated extraordinary capabilities yet remain largely inaccessible at the foundational level. This lack of transparency can hinder the specialized adaptations that geospatial problems demand from real-time disaster response to everyday urban planning challenges.

To clarify the rapidly evolving landscape, Table 1 summarizes the principal technical and licensing characteristics, usage costs, etc., of leading open-source and proprietary LLMs. In this context, general-purpose open-source LLMs such as LLaMA (Touvron et al., 2023), QWen (Bai et al., 2023), and DeepSeek (Guo et al., 2025), among others, have drawn massive attention for their potential adaptability in geospatial innovation. While these models are not explicitly designed for spatial tasks, their open architecture offers flexibility for integration into GIS workflows, enabling researchers to experiment with new techniques in data preprocessing, benchmarking, and machine learning pipelines. Moreover, they also create conditions under which users can rapidly prototype new geospatial methods, integrate advanced spatial indexing, and adapt algorithms to fast-evolving policy or research scenarios. The competitive landscape of LLMs and geospatial AI is expanding rapidly, featuring multiple entrants—each evolving at a pace that underscores the broader "open vs. closed" debate. Indeed, many specialized or hybrid geospatial LLM initiatives have appeared, e.g., ChatGeoAI (Mansourian et al., 2024) and LLM-Find (Ning et al., 2024), reflecting a trend to adapt these Open LLMs for GIScience tasks. In parallel, GIS industry leaders are responding with their own AI strategies: for example, Esri's ArcGIS platform now provides geospatial foundation models (such as "Prithvi" for crop classification and flood segmentation, and "ClimaX" for climate forecasting) that users can fine-tune (Delgadillo et al., 2024), and it even supports the integration of external LLMs (e.g., Mistral (Jiang et al., 2024)) into geospatial workflows. This convergence of open and closed approaches in practice underscores that the future of geospatial AI will be shaped by a blend of community-driven tools and mainstream AI services. Figure 1 contrasts the complementary strengths of open-access and proprietary LLMs, underscoring how transparency and local control on the one hand, and scalability on the other, jointly shape today's GeoAI landscape.

Examining the limitations of proprietary models like ChatGPT is thus not intended merely as a one-to-one comparison but rather as an illustration of the broader issues at stake for GIScience. While ChatGPT's closed infrastructure offers ease of use, it also restricts the ability to validate or refine AI components critical to spatial reasoning. In looking ahead to how GIScience may evolve, we argue that open-source paradigms allow the community to re-engineer AI solutions on their own terms, ensuring reproducibility, compliance with FAIR (Findable, Accessible, Interoperable, and Reusable) principles, and alignment with diverse local contexts. It also allows researchers to investigate emerging trends, such as reinforcement learning (RL) and agent-based modeling for spatial decision-making, in a more transparent setting.

Yet, it is equally important to acknowledge that openness alone does not guarantee widespread acceptance or institutional trust. Reports of government bans on DeepSeek for security or policy reasons highlight that community-driven solutions face scrutiny comparable to (and at times exceeding) their proprietary counterparts. Lawmakers' concerns about data privacy, harmful content generation, or the potential for insecure code can be especially acute for a system in which the source code is open to all. These debates underscore the complexity of forecasting the role of open-source AI in GIScience: Will governmental restrictions limit adoption, or can transparent collaboration and proactive governance ultimately make open platforms more robust and trustworthy?

In this work, we aim to illustrate how open-source AI solutions can address, or at least grapple with, challenges in geospatial intelligence and spark a deeper conversation about the future of GIScience. Our purpose is not to provide a simplistic endorsement but to explore how community-driven architectures might evolve to meet the pressing demands of climate modeling, human-centered GIS, and real-time analytics (Shaw et al., 2016, 2024). The discussion examines the larger horizon for open-source geospatial AI, setting the stage for a truly participatory and innovative GIScience ecosystem.

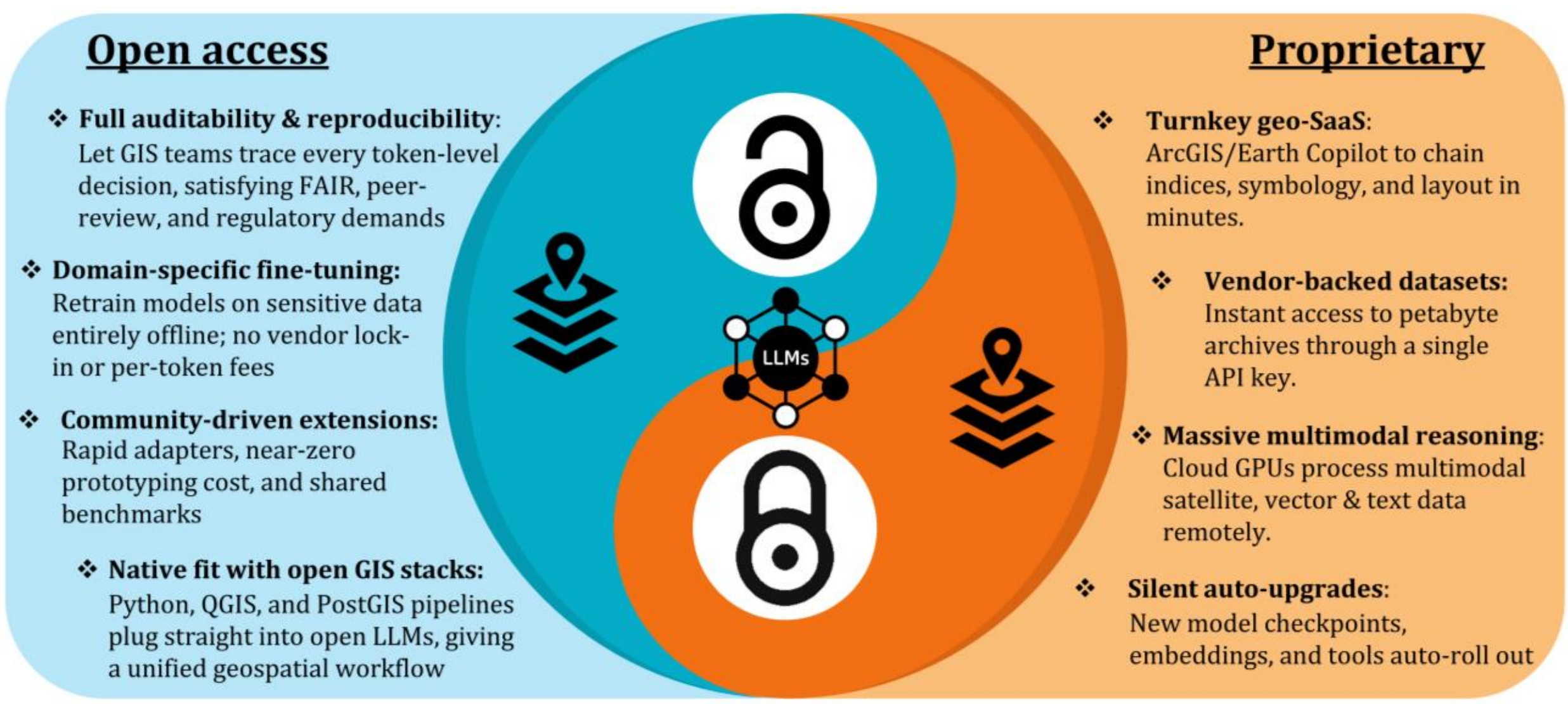

**Figure 1:** Comparative Advantages of Open-Access and Proprietary LLMs in Geospatial AI Workflows.

**Table 1**. Selected open-source and proprietary LLMs.

| Model (release) | Params / Arch. | License / Openness | Highlight Benchmarks | Cost to Use (June 2025) | Principal Backer | GeoAI-relevant / technical notes |
|---|---|---|---|---|---|---|
| **Llama-3 70B** (Apr 2024) | 70 B dense Transformer | Meta Llama 3 Community (open-weight, commercial use) | 82.5 MMLU; top 10 on Chatbot Arena | Free self-host; no official API | Meta (funded internally) | Strong open ecosystem; easy fine-tuning; many geospatial adapters already published |
| **Mistral v2** 140B (Sep 2023) | 140 B MoE (8x17.5B exports) | Apache-2.0 (fully permissive) | 85.7 MMLU; top open MoE model on Arena | Free weights; paid API via Hugging Face Inference | Mistral AI (VC-backed) | 8k context; small memory footprint—good for on-prem GIS pipelines |
| **Qwen 3 235B** (Feb 2024) | Dense & MoE variants: 0.6B–235B total; 128K context | Apache-2.0 (open) | Qwen 3-235B-A22B leads in CodeForces Elo, BFCL, LiveCodeBench; hybrid reasoning mode; multilingual (119 languages) | Free open-weight; API via Alibaba Cloud/HF (pay-as-you-go) | Alibaba Cloud | Supports hybrid reasoning; large context and MoE scaling ideal for multilingual GeoAI corpora |
| **DeepSeek V3** (May 2025) | 671 B total (MoE architecture; 37 B activated at inference) | MIT (code); custom open model license | #1 open model on Arena (Elo > 1300) | Free weights; $ 0.002/1k tokens via NVIDIA NIM micro-service | DeepSeek AI (private Chinese start-up) | Reinforcement-learning-centric training; excels at math/coding → good for spatial tool agents |
| **GPT-4o** (May 2024) | Undisclosed | Closed-source SaaS | Top of MT-Bench; 1st on Arena (proprietary tier) | $5 / 1 M input; $20 / 1 M output tokens | OpenAI (Microsoft $13 B+, Thrive) | Multimodal (text, audio, and vision); strong spatial reasoning but no weights |
| **GPT-o3** (Apr 2025) | Undisclosed | Closed-source SaaS | 87.7 GPQA Diamond; 71.7 SWE-Bench Verified; strong ARC-AGI | $2 in / $8 out per 1 M toks (post-June price cut) | OpenAI (Microsoft, Thrive) | Deep reasoning with image-aware chain-of-thought; promising for multimodal remote sensing analysis |
| **Gemini 2.5 Pro** (Mar 2025) | Undisclosed; 1 M ctx | Closed (Google AI Terms) | 84.1 MMLU-Pro; +39 ELO vs. peers on LMArena | $1.25 in / $5 out per 1 M tokens; free web tier | Google DeepMind (Alphabet) | 1 M-token window & top multilingual scores; excellent for continent-scale GeoAI corpora |
| **Claude 3 Opus** (Mar 2024) | Undisclosed | Closed (Anthropic terms) | Beats GPT-4 (turbo) on many vision and text sets | $15 in / $75 out per 1 M tokens (docs.anthropic.com) | Anthropic (Amazon $4 B, Google $2 B) | Strong safety guardrails; useful for public-facing geo-apps |
| | | | | | | requiring low-risk outputs |
| **Phi-4** (Dec 2024) | 14B dense decoder-only Transformer | MIT (weights & code) | Strong STEM QA performance; top math reasoning benchmarks | Free self-host; Azure/HF inference | Microsoft Research | Compact yet robust reasoning; ideal for math-intensive GIS modeling |

**Notes**: All metrics in the table, including benchmark scores, licensing details, and token-pricing tiers, were captured from the models' public cards, leaderboards, or provider documentation on June 24, 2025. Because these values change rapidly, readers should confirm the latest information directly with the sources before making technical or budgetary decisions. Abbreviations: ctx = context window; MMLU = Massive Multitask Language Understanding benchmark; MoE = Mixture-of-Experts architecture; SaaS = Software-as-a-Service; SWA = Sliding-Window Attention; BFCL = Berkeley Function Calling Leaderboard

## 2. How Open-Source is Transforming GIScience

LLMs are rapidly permeating geospatial workflows, from cleaning raw spatial data to rendering interactive maps, and the choice between open-source and proprietary ecosystems profoundly shapes that trajectory. In the sections that follow, we trace four core stages of the modern GIScience pipeline: 1) data preparation and quality assurance, 2) spatial analysis and modeling, 3) machine-learning–driven inference, and 4) geovisualization and interaction. At each stage, we contrast the freedoms unlocked when model code and weights are released under permissive licenses with the constraints that arise when capabilities are gated behind commercial APIs. Open architectures such as LLaMA-3, Mistral, DeepSeek, K2, and IBM–NASA's Prithvi empower local agencies to retrain, audit, and embed models directly in on-premises tools, while closed platforms like OpenAI's GPT-4/ChatGPT or Google's Gemini offer out-of-the-box scale but at the cost of opacity, recurring fees, and limited customization. By foregrounding these examples, we aim to demonstrate how openness is not merely a philosophical stance but a practical catalyst for reproducible science, regulatory compliance, and place-specific innovation in GeoAI.

### 2.1 Data preparation and quality assurance

Recent studies underscore that LLMs are already capable of routine GIScience "house-keeping," yet the gulf between proprietary and open-access ecosystems remains large. For example, a prototype, named TAGIFY, relies on proprietary GPT-3.5/4 APIs to autogenerate bilingual tags for Estonian open-government data, cutting manual curation time but leaving agencies unable to fine-tune the rules or host the service on-premises, an obstacle under many European data-protection regimes (Kliimask & Nikiforova 2024). By contrast, Memduhoğlu et al. (2024) release both code and weights for a transformer encoder that embeds raw OpenStreetMap (OSM) tags; when those embeddings feed a conventional XGBoost classifier, building-function F1 scores rise by 12% across six cities, and local teams can retrain the model on their own inventories without usage fees. Another example is K2, a 7 billion-parameter Llama-based foundation model for geoscience (Deng et al., 2024). The authors released checkpoints under Apache 2.0 and even ran a public demo from a single RTX 3090 in their university lab; on the GeoBench benchmark, K2 outperforms same-size baselines while remaining fully self-hostable and extensible for downstream spatial tasks (Deng et al., 2024).

These examples suggest a broader pattern: when an LLM's code and weights are openly released, geospatial teams gain full control over the tagging and validation pipeline itself. With TAGIFY, municipal staff must accept whatever heuristics the GPT-4 back-end exposes; if a new domain term appears or a ministry changes its thesaurus, they must submit feature requests to the provider. By contrast, the Memduhoğlu encoder and the K2 checkpoints can be pulled into a local notebook, fine-tuned on fresh OSM exports, and wired into fast gradient-boosting or SQL routines that flag inconsistent, missing, or mistranslated tags in a relatively shorter time. Closed APIs may still offer polished "one-click" hygiene widgets, but the moment a new attribute schema or privacy rule arrives, users must wait for the vendor's next release. Open architectures, therefore, maximize local autonomy and rapid iteration, whereas closed ones trade that flexibility for convenience and vendor support.

## 2.2 Spatial analysis and modeling

LLMs are starting to act as "auto-analysts," chaining together buffers, overlays, network metrics, and even physical simulations from a single prompt. Two development philosophies again diverge on the question of openness. On the closed side, LLM-Geo (Akinboyewa et al., 2025) wires the GPT-4 API into a Python agent that can ingest raw shapefiles, decide which QGIS or GeoPandas tools to call, and return charts or maps without human clicks. In demo case studies, the agent measured wildfire burn severity and calculated school-service catchments in seconds, but every reasoning step flows through OpenAI's proprietary endpoint, so local governments cannot audit the chain-of-thought or adapt it if a new privacy rule blocks cloud uploads. MapGPT (Zhang et al., 2024) pushes the same logic into cartographic design: its LangChain agent exposes 68 specialized "map-element" tools (legend, compass, graticule, etc.) and pipes each call through GPT-4 and DALL·E 3 to generate publication-ready layouts. The consequence is a permanent dependency on the vendor's endpoint, no offline mode, no way to patch tool chains when privacy rules shift, and recurring, token-based API fees.

Projects that rely on open-access LLMs invert those constraints. OmniGeo incorporates a vision encoder and a text backbone into a single multimodal LLM composed of open-access encoders and decoders, which can classify remote-sensing scenes, forecast county-level dementia deaths, and extract street-view perception scores from imagery and metadata (Yuan et al., 2025). The model is trained on 19 public and newly curated datasets spanning health geography, urban function, land-use, scene recognition, and geospatial Named-Entity Recognition (NER). At a finer spatial scale, SpatialRGPT (Cheng et al., 2024) augments an open visual-language model with a relative-depth branch, then distills more than eight million automatically generated 3D QA pairs into spatial-reasoning skill. The proposed model can answer metric queries such as "*How far apart are the pylons*?" as readily as relational prompts like "*Is the crane behind the warehouse?*", and ships with full weights, the data pipeline, and the new SpatialRGPT-Bench evaluation suite, letting practitioners swap in LiDAR depth or rewrite prompts for their own zoning codes.

The contrast is not raw capability but scientific leverage. Closed agents like LLM-Geo and MapGPT deliver turnkey answers yet force users to trust a sealed oracle and budget for perpetual cloud calls. Open-access systems such as OmniGeo and SpatialRGPT demand more engineering, yet they let analysts audit every buffer radius, insert domain-specific error checks, and retrain on local taxonomies, benefits that often outweigh convenience in spatial analysis, where reproducibility, regulatory compliance, and place-specific details are paramount.

## 2.3 Machine learning-driven inferencing

LLMs do not replace traditional geospatial machine-learning (ML) and artificial intelligence (AI) pipelines; they dissolve boundaries between data engineering, model training, and inference.

Whether that integration remains a black-box service or a reproducible research asset depends, again, on openness.

In Esri's ArcGIS Pro 3.4. The Image Analyst extension now ships foundation models, including Microsoft's climate-focused ClimaX and IBM/NASA's flood-segmentation Prithvi-100 m, but only inside an ArcGIS license (Maglia & Wynn, 2024); engineers cannot retrain layers outside Esri's toolbox or benchmark alternate loss functions without waiting for the next release. Google's Vertex AI offers 200 + proprietary and partner LLMs (Gemini 2.5, Imagen 3, Claude, etc.) behind a unified tuning and MLOps console. The service abstracts GPUs and pipelines, yet the resulting checkpoints stay in Google Cloud, and model selection heuristics remain less transparent.

At the other end of the spectrum, the IBM and NASA collaboration on the Prithvi models is a prime example, with their complete model details, code, and data openly available on Hugging Face (https://huggingface.co/ibm-nasa-geospatial). This transparency has empowered researchers to fine-tune these models for specific applications, such as mapping burn scars, on standard computers. This demonstrates that cutting-edge Earth-observation AI can be both powerful and locally reproducible. Google DeepMind's Gemma 3 follows a similar open model approach, offering a lightweight yet sophisticated alternative to the larger Gemini (Team et al., 2025). Its compact size allows a municipal GIS team, for example, to prototype land-cover classifiers directly on a laptop without the need to upload sensitive imagery to the cloud. Community toolkits already make these open-weight models easy to experiment with. A notable example is TorchGeo (via the prithvi-pytorch and TerraTorch extensions), which ships Jupyter notebooks that fine-tune IBM + NASA's Prithvi on EuroSAT and HLS burn-scar tasks (Gomes et al., 2025). This level of transparency and local retraining simply isn't possible with closed, API-only services.

Ultimately, the choice between closed and open AI models in GIScience presents a clear trade-off. Closed systems offer the convenience of immediate scalability and managed infrastructure, but at the cost of locking users into pay-per-use models and obscuring the very model layers that might contain spatial biases. Open-weight models, while demanding more hands-on technical effort, reward users with complete transparency, the ability to tailor models to specific local needs, and adherence to the principles of fair and reproducible science. A hybrid approach appears to be the most sensible path forward. Geospatial professionals can leverage the power of closed platforms for large-scale, on-demand processing while maintaining an open-source, self-hosted model for transparency and reproducibility. This ensures that every environmental forecast, flood map, or land-cover analysis can be thoroughly vetted, independently verified, and, when necessary, openly challenged.

### 2.4 Geovisualization and interaction

A decade ago, GIS visualization meant point-and-click gymnastics in a desktop GUI or hand-written code that chained map-rendering libraries. LLMs are dissolving both barriers by translating natural language instructions into executable cartographic pipelines and by enabling live, conversational interaction with maps. Two paradigms have emerged, each defined by its stance on openness. On the proprietary front, NASA and Microsoft's Earth Copilot (Bryson, 2024) shows what a closed, SaaS-style model can deliver. Powered by GPT-4 on Azure, the chatbot lets researchers ask, "*What was the impact of Hurricane Ian in Sanibel Island?*", then fetches layers from NASA's 100-petabyte archive, composes a multi-panel map, and narrates the trend. Yet every step, from chain-of-thought to rendering code, remains behind an API wall, and users must route queries through Microsoft's cloud licensing stack. Google's Geospatial Reasoning prototype (Schottlander and Shekel, 2025) takes a similar approach: a Gemini agent orchestrates Earth Engine, BigQuery, and Google Maps to answer compound questions related to spatial reasoning and analysis. It may return choropleth layers, heat maps, and plain-language advice, but the core weights, tool-selection heuristics, and color-ramp defaults stay opaque, limiting auditability and customization.

Open-source alternatives invert those constraints. ChatGeoAI (Mansourian & Oucheikh, 2024), built on Llama-2, converts plain sentences into PyQGIS scripts that users can inspect, edit, and rerun locally. Tehran's planners have already retrained the model on Farsi place-names and swapped color palettes without waiting for a vendor roadmap. Likewise, SPOT (Khellaf et al., 2025), an open-source natural-language interface, demonstrates what full transparency can add to map-driven dialogue. SPOT fine-tunes Llama-3 weights so that a plain sentence such as "*Find me a bus platform next to a Cheesecake Factory restaurant and a building with a red roof in Dubrovnik*" is parsed into a YAML scene description, automatically expanded into coherent OSM tag bundles, and executed against an Elasticsearch-backed index. Results are returned on an interactive Leaflet map where each candidate location can be inspected further in Street View, and intermediate files are hosted on GitHub under a permissive license.

These examples point to a clear trade-off. Closed copilots deliver fast, polished outputs that slot neatly into enterprise stacks, but they do so by sealing off the code paths that shape every symbol, color ramp, and query. Open-weight systems take more hands-on effort, yet that effort buys complete auditability and the freedom to tailor cartography to local conventions, accessibility standards, or novel spatial operators. Because every script and model weight are on the table for peer review, visual analytics shift from a vendor-defined black box to a reproducible, community-maintained research tool, one where transparency, not opacity, drives innovation.

## 3 Conclusion and Future Directions

The trajectory of geospatial AI is marked by both remarkable promise and pressing concerns, and open-source platforms illustrate how these tensions may be navigated. On one hand, open frameworks enable greater transparency, reproducibility, and adaptive research, cornerstones of a forward-looking GIScience that seeks to tackle complex problems ranging from environmental sustainability to social equity (Ye et al., 2023). By removing barriers to experimentation, modular solutions can rapidly incorporate emerging methods (e.g., reinforcement learning) and foster global collaboration. The result is a more democratized approach to geospatial intelligence, where practitioners, policymakers, and community stakeholders co-develop and refine cutting-edge tools.

On the other hand, openness introduces challenges that the GIScience community must confront head-on. Recent episodes of heightened regulatory scrutiny, including temporary access restrictions on models such as DeepSeek in certain regions, remind us that AI deployment can become entangled with broader geopolitical, legal, and societal considerations that extend well beyond technical design. Crucially, these developments were not triggered by DeepSeek's open-source status per se; they instead reflect complex political factors. Nonetheless, they underscore a general lesson: the very features that make open-source systems powerful, unfettered access to underlying algorithms and model weights, demand robust governance mechanisms and responsible-innovation practices. These might include robust peer review of code contributions, rigorous security vetting, and frameworks for addressing ethical considerations such as data privacy and algorithmic bias (Huang et al., 2025). For instance, the OECD's AI Principles call for AI that is both innovative and trustworthy, respecting human rights and democratic values (Underwood, 2020); this aligns with the need to ensure accountability and transparency in geospatial AI development. Similarly, the 2025 International AI Safety Report highlights major risks like the leakage of personal information from training data, underscoring why privacy safeguards must be built into AI systems from the start (Gardhouse 2025). The open-source geospatial AI community will need to proactively address the same concerns currently leveled against proprietary LLMs, while leveraging the collective intelligence of its user base to find solutions. In essence, being "open" is not a panacea; active stewardship and adherence to safety and ethics guidelines are required to make open LLM tools truly trustworthy and secure.

Finally, the real opportunity for GIScience lies not in elevating any single platform as a universal remedy, but in cultivating a diverse ecosystem of AI-driven geospatial tools that prioritize transparency, adaptability, and shared progress. In this vision, open-source models serve as testbeds for rapid innovation and domain-specific customization, while commercial services can supply stability, compliance support, and scale often required in production environments. For researchers and policymakers, such pluralism translates into greater freedom to combine and tailor solutions to specific spatial challenges. Framing the conversation around the ecosystem rather than any one model invites ongoing dialogue and collective experimentation. Working together, the GIScience and AI-ethics communities can shape a future in which geospatial AI is simultaneously cutting-edge and trustworthy, advancing the frontiers of spatial research and delivering tangible benefits to society at large under principled oversight.

**Data Availability Statement:**

Data sharing is not applicable to this article as no new data were created or analyzed in this study.

**Disclosure statement:**

The authors employed the OpenAI ChatGPT-4o model exclusively as a language-editing aide to enhance readability, grammar, and stylistic clarity. ChatGPT-4o was not used to create, expand, or alter the scientific ideas, data analyses, or conclusions presented herein, and no AI-generated text has been included verbatim in the manuscript.